\documentclass[12pt,showpacs,showkeys,preprint,prstab]{revtex4}
\newcommand{\unite}[2]{\mbox{$#1\,{\rm #2}$}}

\newcommand{\mean}[1]{\mbox{$\langle{#1}\rangle$}}
\input epsf
\begin{document}

\title{Longitudinal phase space manipulation in energy recovering \\
linac-driven free-electron lasers\thanks{
This work was performed under the auspices of the US-DOE contract \#DE-AC05-84ER40150, the 
Office of Naval Research, the Commonwealth of Virginia, and the Laser Processing Consortium.}}

\author{P.~Piot}
\thanks{Now at Fermilab, Batavia, IL 60510 USA}
\email{piot@fnal.gov}
\author{D.~R.~Douglas}
\email{douglas@jlab.org}
\author{G.~A.~Krafft}
\email{krafft@jlab.org}
\address{Thomas Jefferson National Accelerator Facility, Newport News, VA 23606 USA}
\date{October 9, 2002} 

\begin{abstract}
Energy recovering~\cite{tigner} an electron beam after it has participated in a 
free-electron laser (FEL) interaction can be quite challenging because of the 
substantial FEL-induced energy spread and the energy anti-damping that occurs during 
deceleration. In the Jefferson Lab infrared FEL driver-accelerator,
such an energy recovery scheme was implemented by properly matching the
longitudinal phase space throughout the recirculation transport by employing
the so-called energy compression scheme~\cite{douglas_pac97}. In the present paper,
after presenting a single-particle dynamics approach of the method used to energy-recover
the electron beam, we report on experimental validation of the method obtained
by measurements of the so-called ``compression efficiency" and ``momentum 
compaction" lattice transfer maps at different locations in the recirculation 
transport line. We also compare these measurements with numerical tracking 
simulations. 
\end{abstract}

\maketitle

\section{Introduction}

Free electron lasers (FELs) driven by energy recovery linacs (ERLs)~\cite{tigner}
are now established
as a generic configuration for high average power light sources needed for industrial
applications. In such types of light sources, the ability of a linac-based driver
to provide high electron beam quality (small emittances and high peak current)
is combined with the principal advantage of ERLs: the beam energy is reused.
Because the energy is recovered: (1) the radio-frequency (RF) 
power demand is considerably reduced (the RF-power only provides RF-regulation 
under steady
state operation) and thus the wall-plug efficiency is improved, and
(2) the final beam 
energy can be low and thus issues pertaining to high average current beam dumps 
(dump design, radiological issues,..) are relaxed~\cite{douglas_pac97,neil-prl-2000}.

The operation of ERL-based FELs presents many beam dynamics challenges. These latter are 
principally related to energy jitter induced-instabilities~\cite{merminga_nima2000}, and 
to the proper longitudinal phase space manipulation, the object of the present paper.
During deceleration, the fractional momentum spread of the beam becomes larger due to
anti-damping and if not properly controlled, can yield momentum spread beyond
the acceptance of the downstream beam-line.

A high average power (kW-level) infrared (\( \sim 5 \) $\mu$m) light source, the Ir-Demo,
has recently
concluded operations at Jefferson Lab. In this facility, which has served as a
proof-of-principle 
for an FEL operated using the so-called same-cell energy-recovery (SCER) scheme, careful 
measurement of the longitudinal dynamics were conducted.

In the present paper we address the issues pertaining to the longitudinal phase
space manipulations necessary in such ERL FELs. The paper is organized
as followed: we discuss, in Section 2, general aspects of the longitudinal
beam dynamics in an ERL-driven FEL and introduce the energy compression scheme.
In Section 3, we describe how in practice the longitudinal phase space is manipulated
for the specific case of the Ir-Demo FEL. The experimental characterization
of this ERL light source is presented in Section 4. Finally, we summarize our conclusions in
Section 5.

\section{Longitudinal Dynamics }

\begin{figure}
\centering
\leavevmode
\epsfxsize=160mm
\epsfbox{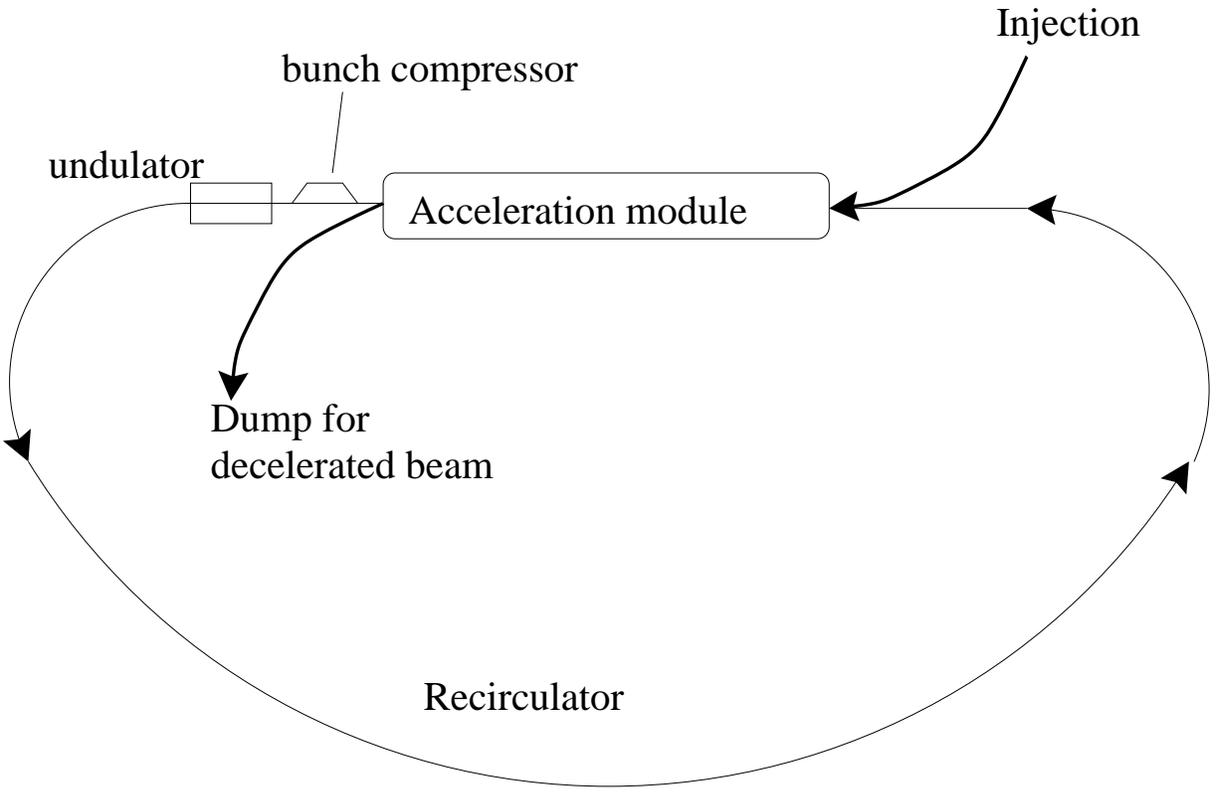}
\caption{\label{fig:erl-fel_generic}Generic configuration for an energy recovering driven free-electron laser.}
\end{figure}

A generic ERL-driven FEL is pictured in Figure~\ref{fig:erl-fel_generic}. The
electron beam,
assumed to be relativistic (so that \( \beta \simeq 1 \) and 
the momentum equals the energy), is injected
in the ERL main linac with an energy of \( \gamma _{o} \), and rms energy
spread \( \sigma _{\gamma _{o}} \). The main linac consists of an RF-accelerating
section  providing an energy gain amplitude \( \gamma _{rf} \) and
operating with a phase \( \varphi _{rf} \) (the phases being referenced to the maximum
acceleration phase). The undulator is located downstream from the linac, after the beam has
been compressed in a magnetic bunch compressor. After participating in
the FEL process, the beam is recirculated and re-injected into the accelerating
section (on the decelerating phase \( \pi -\varphi _{rf} \)) via the so-called
recirculator. The recirculator has several functions:
\begin{enumerate}
\item it bends the beam by a total angle of 360$^{\circ}$, 
\item it provides tuning of the transport path length (to re-inject the beam into
the accelerating section with the proper phase), 
\item it provides variable tuning of bunch time-energy chirp. 
\end{enumerate}
In the Ir-Demo, the decelerated beam is split off the main beam-line
by a dipole and the beam is dumped.

The longitudinal dynamics is given by three requirements: (1) the beam incoming
energy needs to be matched to the FEL resonance condition, (2) the beam must have 
a high peak current at the undulator location, and (3) the recirculator should be 
optimized to minimize the momentum spread after the deceleration so the spent
beam is cleanly dumped.

\subsection{Operation of the free-electron laser}

\paragraph{Energy gain through the RF-accelerating section:}

Downstream from the accelerating section an electron, of coordinate \( s \)
with respect to the bunch centroid, after the accelerating pass
gains the energy: 
\begin{eqnarray}\label{energygain}
\gamma_U (s)= & \gamma_o(s) + \gamma _{rf}\cos (k_{rf}s+\varphi_{rf} )\mbox {.}
\end{eqnarray}
where $k_{rf}$ stands for the RF-field wave vector ($k_{rf}\doteq 2\pi/\lambda_{rf}$, 
$\lambda_{rf}$ being the RF-wavelength). The average beam energy, 
\(\mean{ \gamma _{U}} \), at the undulator location, is: 
\( \mean{\gamma _{U}}=\mean{\gamma _{o}}+\gamma _{rf}\cos \varphi_{rf}  \)

\paragraph{The FEL resonance condition:}

Given the wavelength $\lambda$ at which the FEL must operate, the electron beam
energy must fulfill
the resonance condition: 
\begin{eqnarray}~\label{resonancecond}
\gamma _U=\sqrt{\frac{\lambda _{u}}{2\lambda }(1+K^{2}/2)}
\end{eqnarray}
 where \( \lambda _{u} \) is the undulator period, and \( K \) the undulator
parameter.

\paragraph{The minimum bunch length condition:}

Since the injector cannot directly provide the required high-peak current, it
is necessary to compress the electron bunch by means of magnetic compression.
We characterize the compressor by its first order momentum compaction, \( R^{C}_{56} \),
defined as the linear dependence of the path length though the compressor on
the fractional momentum offset \( \delta  \). Under such a linear assumption,
an electron of initial coordinate \( (s_{i},\delta _{i}) \) is mapped to the
final longitudinal position as: \( s_{f}\simeq s_{i}+R^{C}_{56}\delta _{i} \).
If we assume there is no local momentum spread, the minimum bunch length (\( s_{f}=0 \))
is achieved when the incoming longitudinal position-momentum correlation satisfies 
the relation:
\begin{eqnarray}\label{eqn:longmatch}
\left[\frac{d\delta }{ds}\right]_{s=0}= & -\frac{1}{R^{C}_{56}}\mbox {.}
\end{eqnarray}
 In Eq.~\ref{eqn:longmatch},
we assume the fractional momentum offset is small enough
that second order dependencies on this quantity are insignificant; i.e. we 
assume
\begin{eqnarray} \label{eqn:assumption}
|\delta| \ll \big|\frac{R^{C}_{56}}{T^{C}_{566}}\big| \mbox{,}
\end{eqnarray}
where \( T^{C}_{566} \) denotes
the second order dependence of the path length in the compressor on the fractional
momentum offset. For an achromatic chicane, under the small bending angle 
approximation, we have~\cite{raubenheimer-pac87} $T^C_{566}/R^C_{56} \simeq -3/2$.

Eqs.~\ref{energygain}, \ref{resonancecond} and \ref{eqn:longmatch} yield the 
following values for the required phase and amplitude of the accelerating 
RF-section: 
\begin{eqnarray}
\tan \varphi_{rf} = & \frac{\lambda _{rf}}{2\pi R_{56}}\times \frac{\gamma _{U}}{\gamma _{U}-\gamma _{o}}\mbox
{, and}
\end{eqnarray}
\begin{eqnarray}
\gamma _{rf}=\frac{\gamma _{U}-\gamma _{o}}{\cos \varphi_{rf} } & \mbox {.}
\end{eqnarray}

\subsection{Recirculation}
\paragraph{Longitudinal phase space after the FEL interaction:}
After the electron bunch has contributed to the FEL process, an electron loses an energy 
of $\Delta \Gamma$ on average and the fractional momentum spread of the bunch 
is increased. The average energy downstream from the undulator is: 
$\mean{\gamma_{U+}}=\mean{\gamma_{U}}-\Delta \Gamma$ . 

\paragraph{Longitudinal dynamics in the recirculator:}
As mentioned previously, the recirculator
provides a zeroth order optics knob: it allows one
to tune the path length in such
a way that the beam centroid
is injected in the RF-section with the phase 
\( \psi_{rf} =\pi -\varphi_{rf}  \).
In addition, the recirculator is
used to manipulate the longitudinal phase space, because
it provides \( 360^{\circ } \) bending and generates dispersion (to arbitrary order) 
which can be used to control the linear and quadratic (and higher order) dependence 
of path length on fractional momentum spread in the recirculator. Because the 
incoming momentum 
spread is large, depending on the geometric arrangement of the bending
system, a relation as in Eq.~\ref{eqn:assumption} may not be satisfied,
and hence higher order contributions to the longitudinal transfer map need to
be included. It
is necessary to expand the longitudinal map to the second order in \( \delta  \).
Let \( {\bf x}^{U+}=(x,x';y,y';s,\delta )^{U+} \doteq (x_{1},x_{2},x_{3},x_{4},x_{5},x_{6})^{U+} \)
be the coordinates of the electron at the undulator exit and \( {\bf x}^{R} \), those
after the recirculation, i.e. at the linac entrance. Under the single-particle-dynamics
approximation, the transport from the undulator exit to the linac entrance can be 
modeled by the longitudinal map \( {\mathcal{Z}} \) that is generally 
Taylor-expanded:
\begin{eqnarray}
x^{U+}_k\stackrel{\mathcal{Z}}{\rightarrow }x^{R}_k= \sum_i 
\frac{\partial {\mathcal{Z}}_k}{\partial x^{U+}_i}x^{U+}_i + \frac{1}{2}\sum_i \sum_j 
\frac{\partial^2 {\mathcal{Z}}_k}{\partial x^{U+}_i \partial x^{U+}_j}x^{U+}_i x^{U+}_j 
+ {\cal O}\left(({\bf x}^{U+})^3\right)
\end{eqnarray}
which maps the $k$-th electron coordinate at the undulator exit to the entrance of 
the linac. We now define the first and second order Taylor coefficient (following the
{\sc transport} convention) as: 
$R_{ki}=(\partial {\mathcal{Z}}_k)/(\partial x^{U+}_i)$ and 
$T_{kij}=1/2\times(\partial^2 {\mathcal{Z}}_k)/(\partial x^{U+}_i \partial x^{U+}_j)$. 
Using these definitions and only considering the $s$-coordinate we have:
\begin{eqnarray}~\label{afterrecirc}
s^{U+}\stackrel{\mathcal{Z}}{\rightarrow }s^{R}=(R_{55}+\sum_{j=1}^{4}T_{55j}x_j^{U+})s^{U+}
 + T_{555}(s^{U+})^{2}+(R_{56}+\sum _{j=1}^5 T_{56j}x_j^{U+})\delta^{U+}+T_{566}
(\delta ^{U})^{2}+...\mbox{.}
\end{eqnarray}

\begin{figure}
\centering
\leavevmode
\epsfxsize=160mm
\epsfbox{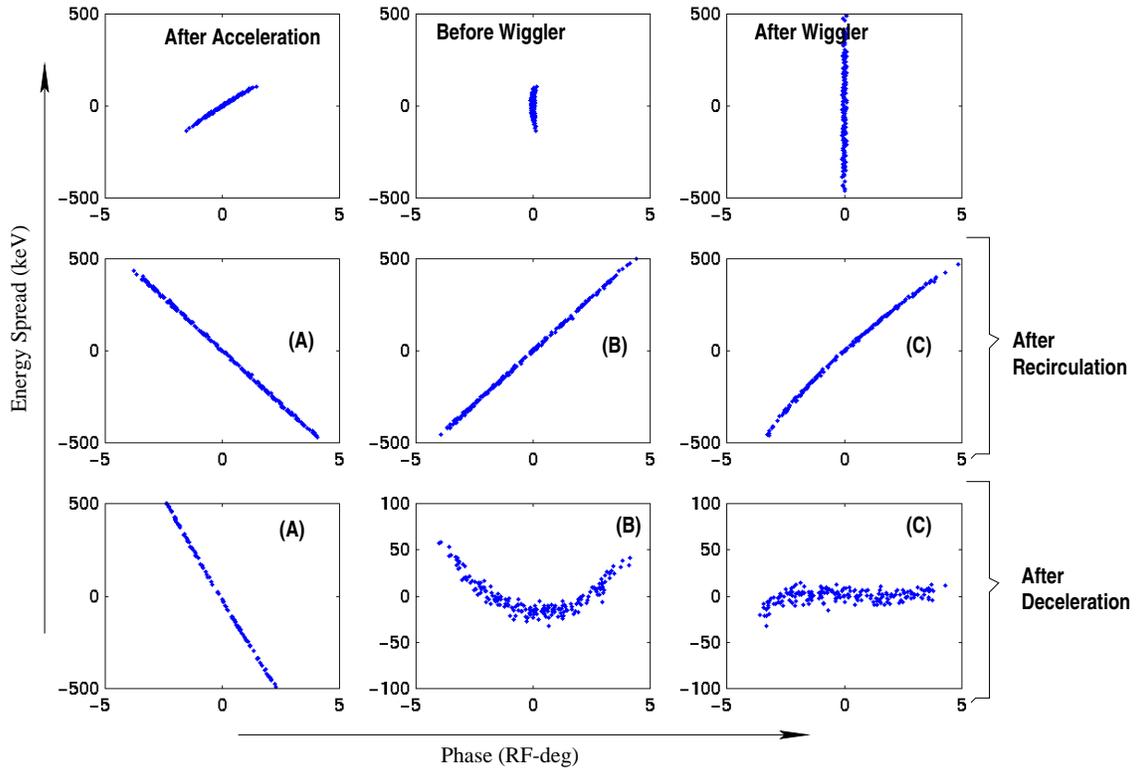}
\caption{\label{fig:nrjrecov_scheme} 
Energy compression scheme: The first row (from left to right) presents
the longitudinal phase space at the linac exit, after the compression chicane,
and just after the wiggler interaction has taken place; the second row 
show longitudinal phase space at the entrance of the linac just prior 
to deceleration for three different choice of $R_{56}$ and $T_{566}$ (
for ({\bf A}) -0.2 and 0. m, for ({\bf B}) 0.2 and 0 m and for ({\bf C}) 
0.2 and 3.0 m). The result for the three cases after deceleration are shown 
in the third row.}
\end{figure}

\paragraph{Deceleration for energy recovery:}
As the beam is decelerated in the RF-section, an electron of coordinate $s'$ downstream 
of the recirculator will have the energy  
\begin{eqnarray}
\gamma _{D}(s')=\gamma_{R}(s')-\gamma _{rf} \cos (k_{rf}s'+\psi_{rf})  & \mbox {,}
\end{eqnarray}
downstream of the RF-section. Here $\psi_{rf}=\pi-\phi_{rf}$ and $\gamma_{R}$ is the 
electron energy before deceleration (i.e. at the recirculator exit). From this latter 
equation the average energy takes the form: 
\( \mean{\gamma _D}=\mean{\gamma _{o}}-\Delta \Gamma \). The quantity of interest 
after deceleration is the fractional momentum offset, 
$\delta^{D}(s')=\gamma_D(s')/\mean{\gamma_D}-1$;
\begin{eqnarray}~\label{matchdecel}
\delta^D(s')=\frac{\gamma_{rf} \left[(k_{rf}s')^2/2\cos(\psi_{rf})+k_{rf}s'
\sin(\psi_{rf})\right]}{\mean{\gamma_{U+}}-\gamma_{rf} \cos(\psi_{rf})} + 
\delta^R(s')\frac{\mean{\gamma_{U+}}}{\mean{\gamma_{U+}}-\gamma_{rf}\cos(\psi_{rf})} 
&\mbox{,}
\end{eqnarray}
with $\delta^R(s')$ being the fractional momentum offset upstream from the RF-section
before the second pass. 
In Eq.~\ref{matchdecel}, we have expanded the fractional momentum offset 
up to the second order in $s'$ because the incoming bunch length, $\sigma_{s'}$, does 
not a priori satisfies the condition $\sigma_{s'} \ll 1/k_{rf}$.  
Eqs.~\ref{afterrecirc} and \ref{matchdecel} suggest a mechanism 
for counteracting the momentum spread generated during deceleration. The method 
consists of setting up the recirculator in a way to impart both a linear 
and quadratic a position-energy correlation, and allowing the bunch to decompress
before the decelerating pass. Formally one wants to match the linear
 and quadratic dependence of 
$\delta^R$ to
\begin{eqnarray}
\left[\frac{d \delta^D}{ds'}\right]_{s'=0}=-\frac{\mean{\gamma_{rf}}}
{\mean{\gamma_{U+}}} k_{rf} \sin(\psi_{rf})
\mbox{, and}
\left[\frac{d^2 \delta^D}{d^2s'}\right]_{s'=0}=-\frac{\mean{\gamma_{rf}}}
{2 \mean{\gamma_{U+}}} k^2_{rf} \cos(\psi_{rf})
\end{eqnarray}
where the latter equation insures the RF-induced curvature is canceled. 
The technique is illustrated in Figure~\ref{fig:nrjrecov_scheme}. In this 
figure, the last row contains the longitudinal phase spaces downstream
from the linac after deceleration for three different choices of recirculator
optics. Plot (A) illustrates that the phase space slope should be
properly chosen, as otherwise excessive energy spread results. Once the
slope is properly chosen, plot 
(B) of this last row illustrates the importance of RF-induced curvature as an 
lower limit for momentum spread. In plot (C) we show how, using the 
recirculator to impart a quadratic dependence of the fractional momentum spread 
on the longitudinal position upstream from the linac, allows the compensation 
of RF-induced curvature which, in turn, greatly reduces the fractional momentum 
spread. In this latter case the fractional momentum spread is limited 
by a 3rd order aberration (as seen by the ``S" shape of the phase space). 

\section{Example of the Ir-Demo}
\begin{figure}
\centering
\leavevmode
\epsfxsize=160mm
\epsfbox{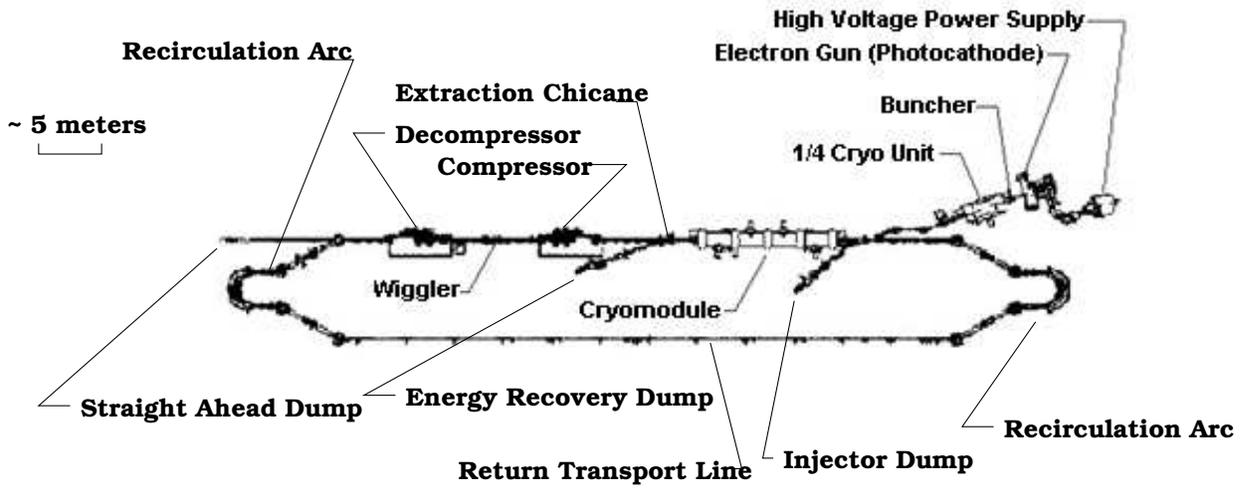}
\caption{\label{fig:irdemo_overview}Overview of the Ir-Demo free-electron laser of Jefferson Lab.}
\end{figure}

In the Ir-Demo (see top view in Figure~\ref{fig:irdemo_overview}), the electron beam, 
is generated by a~\unite{350}{keV}~photoemission electron gun~\cite{engwall-pac97}, 
and accelerated~\cite{piot_epac1998} to $\sim$10~MeV  by two 
superconducting radio-frequency (SRF) CEBAF-type cavities 
(5-cell $\pi$-mode standing-wave cavities largely of the
same design as those in the CEBAF accelerator~\cite{arnps}) mounted as a 
pair in the so-called ``quarter cryounit". The beam is then injected into the main 
linac which is composed of one cryomodule, containing 8~CEBAF-type SRF cavities. 
The linac can provide a net energy gain of approximately \unite{37}{MeV} (but is 
operated to provide 28~MeV energy gain for the results presented in this paper). 
The operating frequency of the RF-system is 1.497~GHz, so 
$\lambda_{rf}\simeq 0.20026$~m. 
This cryomodule is followed by two 4-bend achromatic chicanes that bypass the 
FEL resonator mirrors and provide longitudinal phase space manipulation. 
The first 
chicane, upstream from the undulator, serves as bunch compressor and the second
chicane naturally decompresses the upright bunch. The undulator 
is located between the two chicanes. Soon after the downstream chicane
the beam is recirculated, by the means of 
a recirculator with variable (linear and quadratic) momentum compaction and path 
length, back to the entrance of the cryomodule. The path length is chosen so
the second pass bunches have the proper time of arrival: the electron bunches are on the 
decelerating phase of the radio-frequency wave. The electrons are decelerated down to 
10~MeV and are separated from the 48~MeV beam and dumped in the ``energy recovery 
dump" by the means of the ``extraction chicane". \\

A complete description of the driver-accelerator can be found in 
References~\cite{douglas_pac97,douglas-linac2000}.
In terms of the longitudinal beam dynamics, the first matching point is at the 
undulator location where a longitudinal
waist (minimum bunch length) is required. Given the 
longitudinal phase space at the injector front-end, the SRF-linac settings must 
be tuned in amplitude and in phase for achieving the desired longitudinal phase 
space correlation, \( \simeq d\delta /ds \), to match the momentum compaction 
\( R_{56}^{C}\simeq -28.8 \)~cm of the first chicane according to
Eq.~\ref{eqn:longmatch}. 
In the nominal operating conditions, the aforementioned requirements result
in operating the linac $\varphi_{rf}=$\unite{-10}{^{\circ}} off crest for an accelerating 
voltage of approximately $\gamma_{rf}m_ec^2=$\unite{37}{MeV}
to provide an FEL wavelength $\lambda\simeq$\unite{5} {\mu m}.
Downstream from the undulator the beam-line consists of (1) a magnetic chicane 
similar to the one upstream but which now acts as a bunch decompressor, and (2) a 
recirculation loop. The recirculation loop incorporates two 180~$^{\circ}$ arcs linked  
by a straight line section, the ``return transport line", which consists of 
six FODO cells having a $90^{\circ}$ betatron phase advance per cell. 
The arcs are based on the MIT-Bates accelerator design~\cite{flanz-nima85}; they 
each provide a total bending angle of \unite{180}{^{\circ}}. They include four 
wedge-type dipoles, each bending the beam by an angle of about \unite{\pm 28}{^{\circ}} 
alternatively, installed in pairs symmetrically around a \unite{180}{^{\circ}} dipole. 
In addition to providing the desired bending, the arcs are also used to adjust the 
total beam path length of the recirculated beam. 
For such purposes, the arc is instrumented with a pair of horizontal steerers 
located upstream and downstream the 180~$^{\circ}$ dipole to vary the reference orbit path 
length inside the 180~$^{\circ}$ magnet. This provides a path length adjustment of the 
order of $\pm \lambda_{rf}/2$ off the nominal length ($\simeq 501.5\times \lambda_{rf}$). 

\begin{figure}
\centering
\leavevmode
\epsfxsize=160mm
\epsfbox{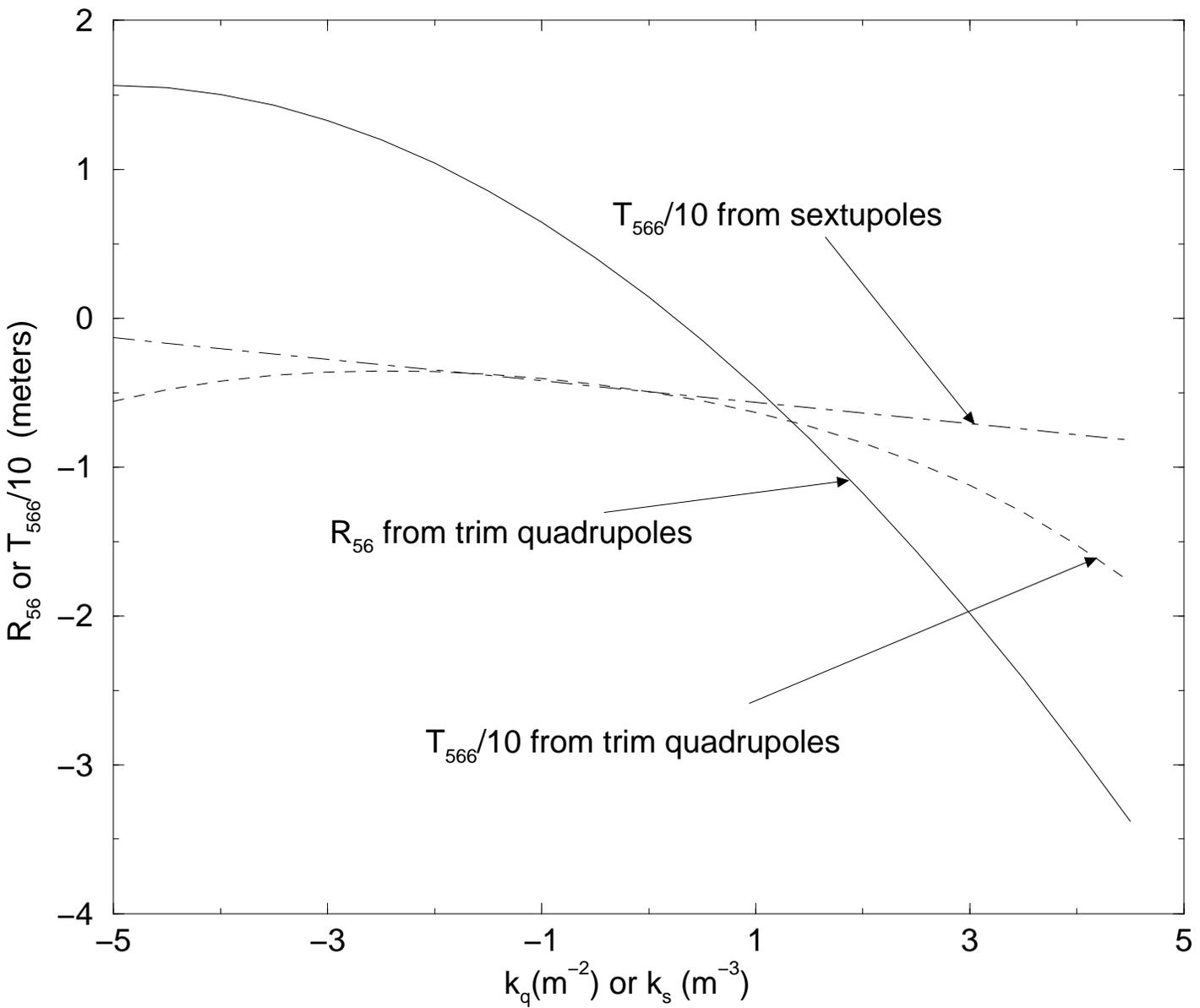}
\caption{First ($R_{56}$) and second ($T_{566}$) order momentum compaction evolution, 
for one arc of the recirculator, versus the settings of the trim quadupoles and 
sextupoles. \label{fig:R56andT566}}
\end{figure}

Two families of quadrupoles (the trim quadrupoles) and sextupoles are 
installed in the arc to provide both linear and quadratic energy dependent path 
length variation that are necessary in the ``energy-compression" scheme needed to 
properly energy recover the beam~\cite{douglas-tn98-025}. 
The quadrupoles act both on the linear and quadratic momentum compaction of the 
recirculator (these quantities are henceforth noted $R^{U+\rightarrow R}_{56}$ and 
$T^{U+\rightarrow R}_{566}$) while the sextupoles only impact the quadratic momentum 
compaction; the magnitude of the impact of these elements on the momentum compaction 
is illustrated in Figure~\ref{fig:R56andT566}. 
When the quadrupoles and sextupoles are unexcited, the arcs are operated in a 
non-isochronous mode ($R^A_{56}=13.1$~cm for a single 180$^\circ$~arc). However under nominal operation, i.e. 
when the FEL is operating and the linac is in energy 
recovery mode, because of the need for energy compression, the sextupoles and 
quadrupoles of one family are excited to proper values in order to provide the 
required parameters. The momentum compaction of the recirculation loop is related 
to those of the individual components via:
\begin{eqnarray}
R_{56}^{U+\rightarrow R}=R_{56}^{C}+R_{56}^{A1}+R_{56}^{A2} \mbox{,}
\end{eqnarray}
where the subscript $C$, $A1$ and $A2$ indicate the quantity corresponds 
respectively to the ``decompressor" chicane, the first and second arc. Similar 
relations yield for the second order momentum compaction. 

\section{experiment in the Ir-Demo}

\subsection{Measurement of the initial conditions at the undulator}

\begin{figure}
\centering
\leavevmode
\epsfxsize=160mm
\epsfbox{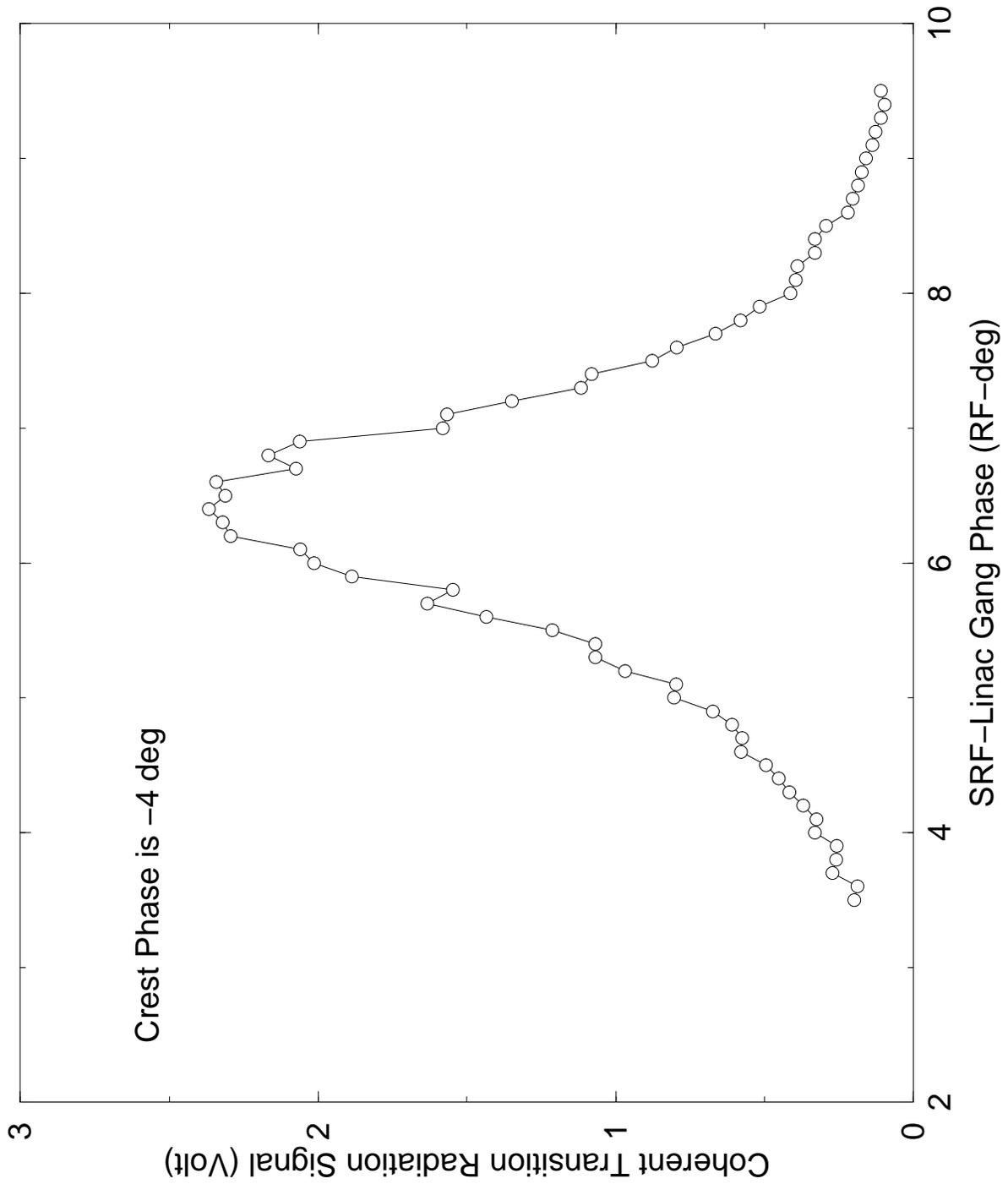}
\caption{CTR signal versus SRF-linac phase. The maximum CTR signal coincides with  
the shortest achieved bunch length. For the corresponding phase, the longitudinal 
phase space is up-right.\label{fig:ctrvsgang}}
\end{figure}

Experimentally, the accelerator module amplitude is set up accordingly to numerical
simulation, then the phase is tuned to obtain the minimum bunch length at the
undulator (i.e. the highest peak current). The bunch length is 
monitored~\cite{krafft-epac1998} by
detecting the coherent transition radiation (CTR) emitted in the backward direction
as the electron bunch crosses an aluminum foil adjacent to the undulator. The power 
density radiated by a bunch of $N$ electrons is 
\begin{eqnarray}
[\frac{dP}{d\omega d\Omega }]_{N}=[\frac{dP}{d\omega d\Omega }]_{1}\times (N+N(N-1)
|\int ^{+\infty }_{-\infty }dtS(t)\exp -i\omega t|^{2}) & \mbox {,}
\end{eqnarray}
where $[\frac{dP}{d\omega d\Omega }]_1$ is the single electron power density. Thus 
since the Fourier transform of a bunch with characteristic length
$\sigma_s$ extends to frequency $\omega\sim c/\sigma_s$, detecting the CTR at 
frequencies close to this frequency provides indirect information on the bunch 
length. An example of such a measurement is presented in Figure~\ref{fig:ctrvsgang}.

\begin{figure}
\centering
\leavevmode
\epsfxsize=150mm
\epsfbox{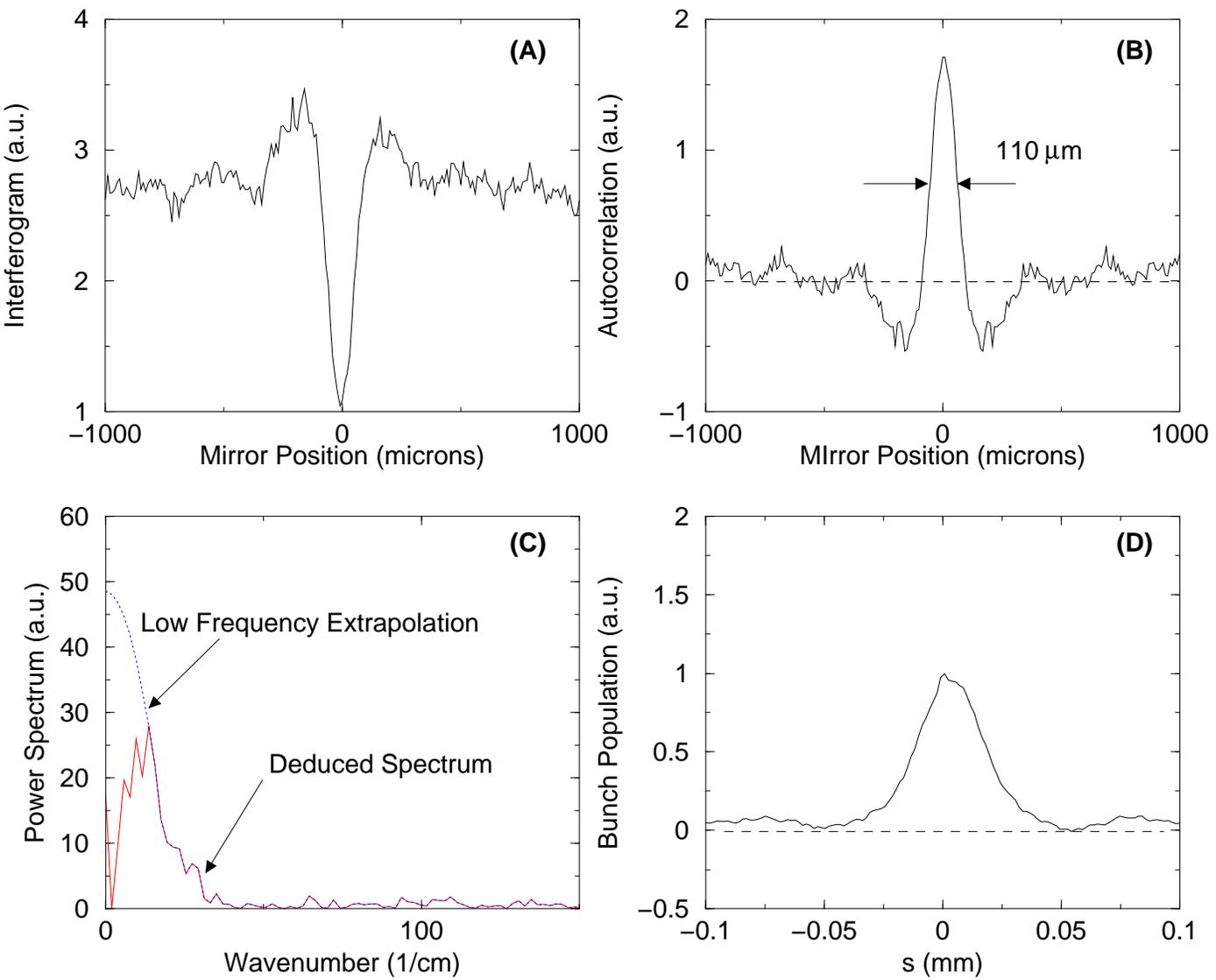}
\caption{Example of bunch length measurement. 
\textbf{(A)} raw data obtained from the Michelson polarizing interferometer, 
\textbf{(B)} deduced autocorrelation from the interferogram, 
\textbf{(C)} power spectrum obtained by Fourier-transforming the autocorrelation, 
\textbf{(D)} charge density longitudinal distribution obtained by applying the 
Kroenig-Kramer relations on the power spectrum to recover the (missing) phase 
information.\label{fig:bunchlength}}
\end{figure}

Once the phase is properly tuned, the bunch length is measured by performing
Michelson interferometry~\cite{happek-prl,piot-tn} of the CTR signal, since the 
radiation pulse emitted
by the electron bunch mirrors the electron distribution. From the measured interferogram,
one can deduce the autocorrelation function and (in virtue of the Wiener Kintchine
theorem) the power spectrum of the CTR. A logarithmic 
Hilbert-transform~\cite{burge-Lond76,lai-pre95} of
the latter power spectrum allows one to reconstruct the electron longitudinal charge
density. A typical measurement of an interferogram along with the subsequent
calculations is shown in Figure~\ref{fig:bunchlength}. Routinely, the bunch length 
achieved in the Ir-Demo FEL is approximately \( 100 \)~\( \mu  \)m (rms) for a bunch 
charge of 60~pC.

\begin{figure}
\centering
\leavevmode
\epsfxsize=160mm
\epsfbox{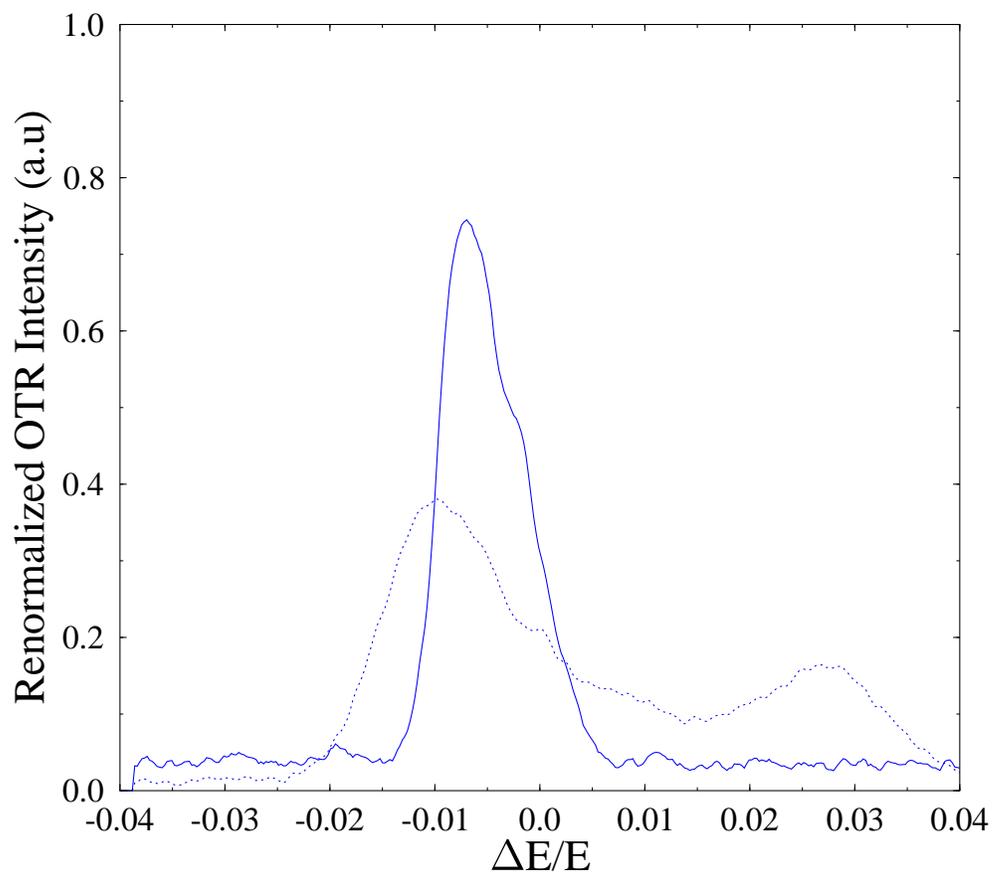}
\caption{Impact of the FEL process on the beam fractional momentum profile. 
The energy profiles are measured in the "decompressor" symmetry point.\label{fig:felonoff_dpp}}
\end{figure}

The fractional momentum spread of the beam is deduced from a profile measured at 
the high dispersion point of the decompressor chicane; the three quadrupoles 
upstream from the chicane are used to insure the betatron contribution to the 
beam profile is insignificant.
In Figure~\ref{fig:felonoff_dpp} we compare the fractional momentum profiles with 
and without operating the FEL (the FEL process can be turned on/off by 
tuning/detuning the length of the optical resonator). The typical rms fractional 
momentum spread before the FEL interaction is about 0.4\%, which grows to 
to 1.2\% downstream from the FEL when the laser operates. 

At the undulator location, the bunch length has been minimized, the 
longitudinal phase space is thus expected to be up-right; i.e. no 
position-energy correlation exists.

\subsection{Experimental setup to measure longitudinal transfer maps}
The measurements of the longitudinal transfer map ${\cal Z}$ is very 
difficult. Instead, we perform a perturbative measurement which provides 
information on the Taylor expansion of the map. The technique consists of 
perturbing the initial conditions at a given location $i$ (initial RF-phase, $\phi^i=2\pi s^i/\lambda_{rf}$, 
or initial energy $\delta^i$) and measuring the relative
time-of-flight (TOF) to a 
downstream position $f$ (the time-of-flight is also measured in RF-phase units). 
The two aforementioned excitations allow the perturbative measurement of 
the $\partial s^f/\partial s^i$ or $\partial s^f/\partial \delta^i$ expansions of 
the map which gives (see Eq.~\ref{afterrecirc}):
\begin{eqnarray}~\label{eqn:tmap-phi-phi}
\frac{\partial s^f}{\partial s^i}=
R_{55}^{f\rightarrow i}+\sum_j T_{55j}^{f\rightarrow i} x_j + 
2 T_{555}^{f\rightarrow i} s^i \mbox{ and,}
\end{eqnarray}
\begin{eqnarray}~\label{eqn:tmap-delta-phi}
\frac{\partial s^f}{\partial \delta^i}=
R_{56}^{f\rightarrow i}+\sum_j T_{56j}^{f\rightarrow i} 
x_j + 2 T_{566}^{f\rightarrow i} \delta^i \mbox{.}
\end{eqnarray}
Henceforth we will term $\partial s^f/\partial s^i$ and 
$\partial s^f/\partial \delta^i$ respectively as 
``compression efficiency" and ``momentum compaction" maps, and we will instead 
work in RF-phase unit: 
$\partial s^f/\partial s^i \rightarrow \partial \phi^{f}/\partial \phi ^{i} $ and 
$\partial s^f/\partial \delta^i \rightarrow \partial \phi^{f}/\partial \delta ^{i}$.
In essence, the measurement of the compression efficiency or momentum compaction 
maps reduces to a relative TOF variation measurement.

Measurement of TOF is performed by detecting the phase of a signal produced by
the TM\( _{010} \) waves excited as the electron bunches traverse a resonant 
1.497 GHz stainless
steel cavity~\cite{krafft_aip_95,piot_epac2000,hardy-pac97}. The principle of the
TOF measurement is to measure the phase of
the beam induced voltage since it has constant phase with respect to the
electron bunches. The phase of the rf
signal coming from the cavity is mixed
with the reference signal, which may be phase shifted by means of a programmable phase
shifter. The mixer 
output, after removal of high frequency component with a low pass filter, is
calibrated by performing a procedure that consists of varying the phase shifter
to find the output maxima from the mixer.
Once the measurement is calibrated, the phase shifter phase 
is set so that for the nominal conditions of the machine the cavity is operated
at zero-crossing of the mixer.
Thus a change in TOF give rise to a
linear change in the mixer output.

The change in the TOF induced by perturbing the beam initial phase or energy 
condition upstream can provide  \( \partial \phi ^{f}/\partial \phi ^{i} \) or 
\( \partial \phi ^{f}/\partial \delta ^{i} \) respectively. In the Ir-Demo the 
former kind of measurement is performed by varying the phase of the photo-cathode 
drive laser (with respect to the RF-master oscillator) whereas the latter type of 
measurement is done by modulating the gradient of the last SRF-cavity of the 
cryomodule.  In the accelerator three pickup cavities have been installed. Their 
locations are downstream of: (1) the crymomdule (C1), the (2) first 180~$^{\circ}$
(C2) and (3) second 
180~$^{\circ}$ arc (C3).
To expedite the measurements, the quantity varied (i.e. laser phase for 
\( \partial \phi ^{f}/\partial \phi ^{i} \)
transfer map and cavity gradient for \( \partial \phi ^{f}/\partial \delta ^{i} \)
transfer map), is varied at 70~Hz during the acquisition of measurement
data.

\subsection{Numerical model for calculation of longitudinal transfer map}

The measurement of the compression efficiency map provides important
information on the performance of the bunching process and can give some insights
on the bunch length. Because the map is measured between the photocathode and
the pickup cavities, it cannot be simulated using standard single particle dynamics 
codes, but needs to be computed using a particle tracking code, e.g.  
\textsc{parmela}~\cite{parmela_refer}, 
which includes non relativistic effects such as phase slippage effects in accelerating
cavities. The technique we have used to compare the measurements with numerical 
simulations is as follows: 
we use \textsc{parmela} to generate uniform macroparticles distribution
over a given extent in RF-phase (or time) at the photocathode surface. The corresponding
phase of emission \( \phi _{k}^{i} \) of the $k$-th macroparticle at the photocathode
surface is recorded and the macroparticles populating this uniform distribution
are tracked along the beam-line. For the tracking the space charge subroutine
is not activated, and each macroparticle is assumed to be the bunch centroid of
bunches emitted at different drive-laser phase, we then compute the phase of
arrival \( \phi _{k}^{f} \) at the desired pickup cavities. The couple (\( \{\phi _{k}^{i} \),
\( \phi _{k}^{f}\}_{k=1,...,N} \)) directly gives the phase-phase transfer
map which can be compared to the experimental data. 

In order to generate energy-phase
transfer maps, we use the arbitrarily high order code \textsc{TLie}~\cite{tlie-refer}
based on a symplectic integrator: the energy offsets achieved when modulating
the gradient of the last cryomodule cavity are directly used by the code to calculate
the TOF up to the desired the pickup cavity. The couple (\( \{\delta _{k}^{i} \),
\( \phi _{k}^{f}\}_{k=1,...,N} \)) provides the energy-phase correlation and
again can be compared with the data.

\subsection{Path length adjustment}
In passing we note that one of the installed detectors
can also be used to precisely setup the path length 
of the recirculator; in such a case the detector C1 is used. The beam is first 
dumped in the ``straight ahead dump" (see Fig.~\ref{fig:irdemo_overview}) and the 
phase of the reference signal used for C1 is shifted so that the signal at the 
mixer output is maximized. The beam is then recirculated and the path length 
is varied using the dedicated horizontal steerers located at both entrance and 
exit of the two 180$^{\circ}$ bends. The path length is optimized when the 
signal measured at C1 is zeroed: this corresponds to the case when the 
beam induced voltage generated by the first pass and recirculated beams exactly 
cancel.

\subsection{Experiments and simulations}
The compression efficiency and momentum compaction transfer map measurement 
have been extensively used during the commissioning of the Ir-Demo, to verify 
our model, but also in routine operation, to insure the accelerator, and
especially 
the recirculator loop, is properly set up.
\subsubsection{Compression efficiency}
\begin{figure}
\centering
\leavevmode
\epsfxsize=160mm
\epsfbox{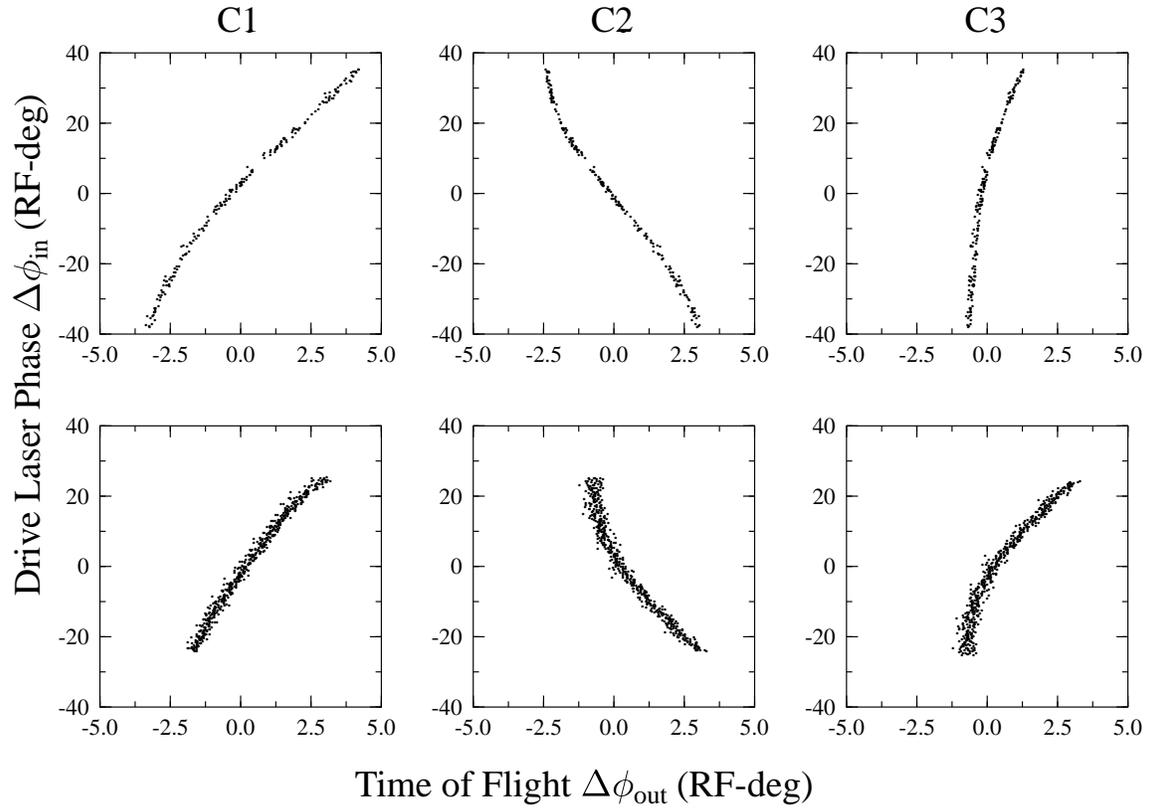}
\caption{Comparison of the phase-phase beam transfer function between 
the photocathode surface and the three different pickup cavities 
(pickup C1, C2, and C3) {\bf (bottom row)} with the one simulated 
using {\sc parmela} {\bf (top row)}.\label{fig:phi-phi-tmap}}
\end{figure}

In Figure~\ref{fig:phi-phi-tmap}, we compare a typical measurements of the 
\( \partial \phi ^{f}/\partial \phi ^{i} \) maps at the three different 
detectors  (C1, C2, and C3) with the maps generated via simulations. 
We generally observe a good agreement with between the measurement and 
the expectations. The slight disagreement, e.g. as the one observed at 
detector C3, is attributed to mis-steering in the arc~2 which makes larger the 
contribution of the $T_{55j}$ terms (with $j=1,...,4$) in 
Eq.~\ref{eqn:tmap-phi-phi}. The second arc transport is more vulnerable to 
such misalignments since the bunch length is larger compared to arc~1. \\

More quantitative information can be obtained by performing a nonlinear fit 
of the transfer map presented in the Figure. This can give some insight on 
the linear \( R_{55} \) and quadratic, \( T_{555} \) compression efficiency 
coefficients between the photocathode and the pickup cavities: the results are 
presented in Table~\ref{tab:m55t555}. \\

\begin{figure}
\centering
\leavevmode
\epsfxsize=160mm
\epsfbox{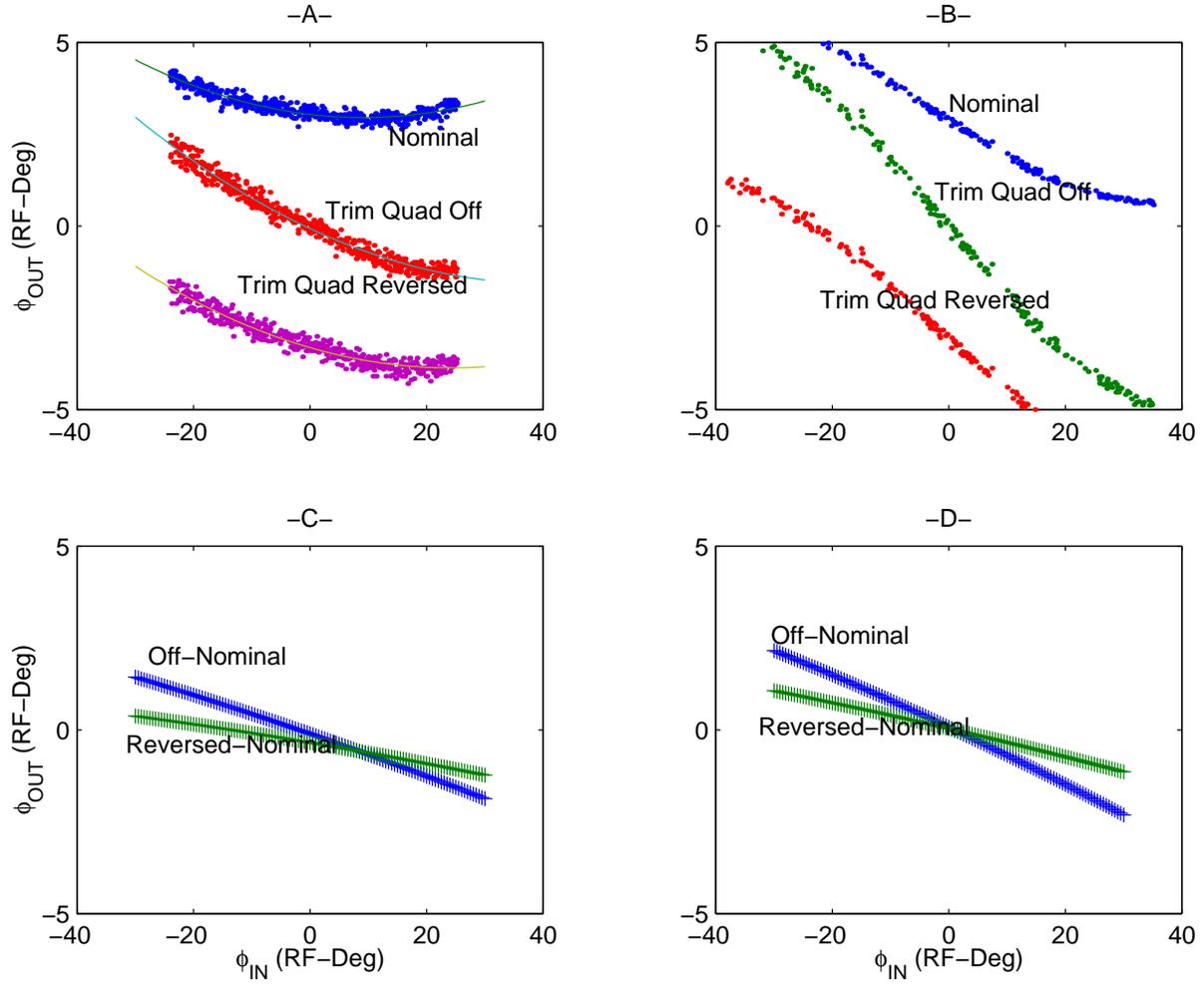}
\caption{Example of a difference compression efficiency transfer map measurement 
at the detector C2. The map is measured {\bf (A)} and simulated {\bf (B)} for 
three excitations of one pairs of trim quadrupole in the arc~1 (quadrupoles 
set up to their nominal values, turned off, and set up to a value opposite to 
their nominal value). Though the simulated and measured maps differ, the
measured {\bf (C)} and {\bf (D)} simulated difference maps results are in 
better agreement.\label{fig:do-55-quad}}
\end{figure}

The compression efficiency map measurement was also used as a trouble shooting 
tool. In Figure~\ref{fig:do-55-quad} we present a series of measurement we 
performed using the detector C2: the compression efficiency was measured for 
three settings of one of the quadrupole pairs of arc~1: 
(a) the quadrupoles are excited to their nominal 
values, (b) they are turned off, and (c) they are powered to a value opposite to 
case (a). The comparison of the measured transfer map with the simulated one 
shows some disagreement (at that time we could not accurately know precisely
the machine settings of the upstream beam-line, e.g., in the 
injector). However, if one compares the difference 
measurement (i.e. calculated as (b)-(a) and (b)-(c)) with the corresponding 
difference simulations, the agreement becomes very good. Such an example 
illustrates how a measurement of transfer maps for various perturbations of the 
lattice (here using quadrupole pairs) can be used to verify that the different 
lattice elements that compose the beam-line perform as expected. This method
was in fact used to diagnose a reinjection phase error during commissioning~\cite{douglas-sum}.

\subsubsection{Momentum compaction:}
The momentum compaction transfer map was measured by modulating the gradient 
of the last cryomodule cavity by $\pm 1$\%. \\

\begin{figure}
\centering
\leavevmode
\epsfxsize=160mm
\epsfbox{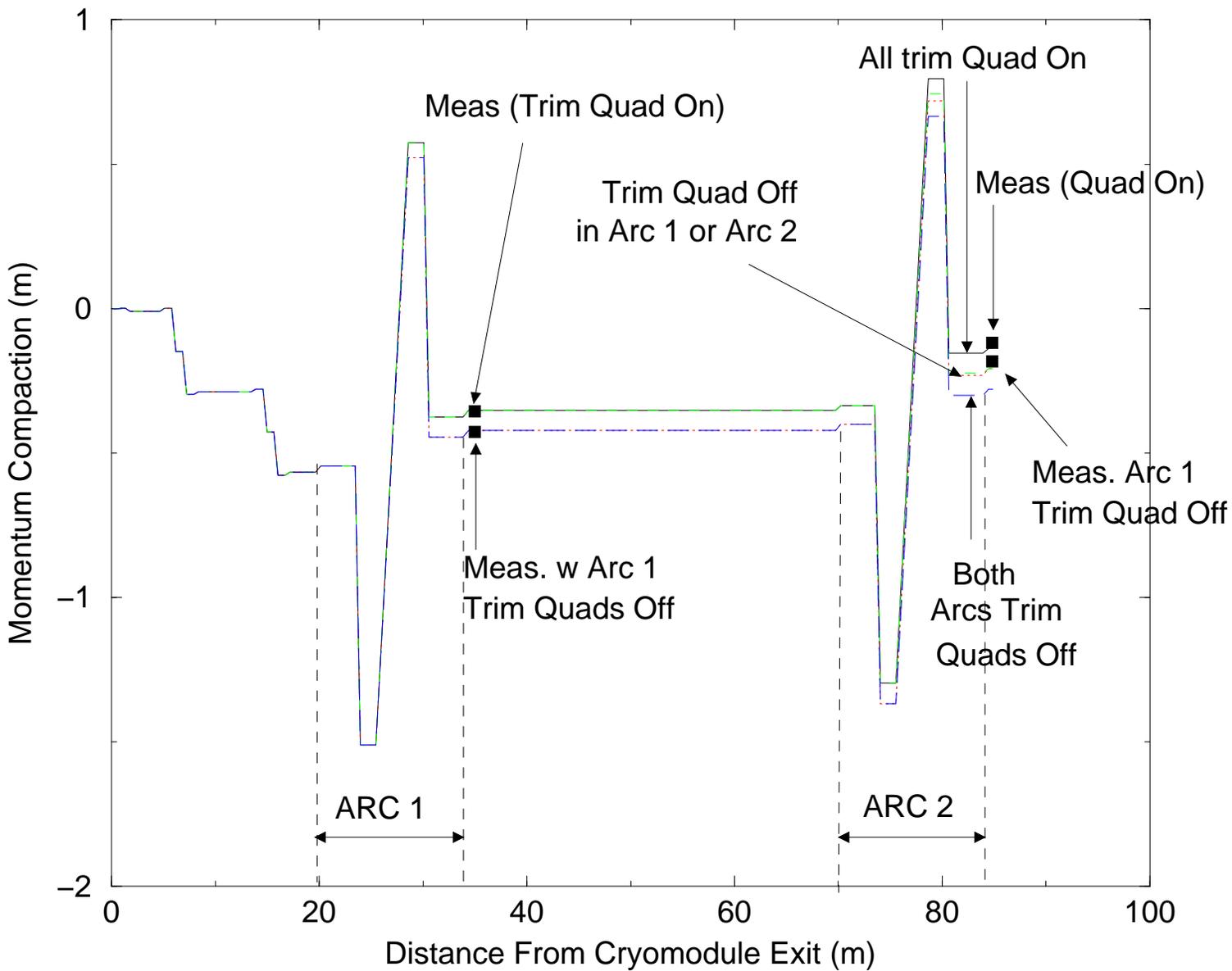}
\caption{Evolution of the linear momentum compaction along the beam-line, from the 
linac exit to the linac entrance.\label{fig:r56-evolution}}
\end{figure}

The linear expansion of the transfer map, $R_{56}$, was measured at both 
C2 and C3. In Figure~\ref{fig:r56-evolution}, we compare the expected evolution 
of the linear momentum compaction for different excitations of the ``trim 
quadrupoles" with the measurement obtained at C2 and C3: the agreement is 
excellent. A more quantitative measurement was performed using C2, the momentum 
compaction from the cryomodule exit up to C2 was measured as a function of the 
trim quadrupole settings, the comparison of such a measurement with the expected
values obtained with the second order optics code 
\textsc{dimad}~\cite{dimad-refer}, again the good agreement confirms the 
usefulness of time-of-flight diagnostics for setting up the recirculation.\\

\begin{figure}
\centering
\leavevmode
\epsfxsize=160mm
\epsfbox{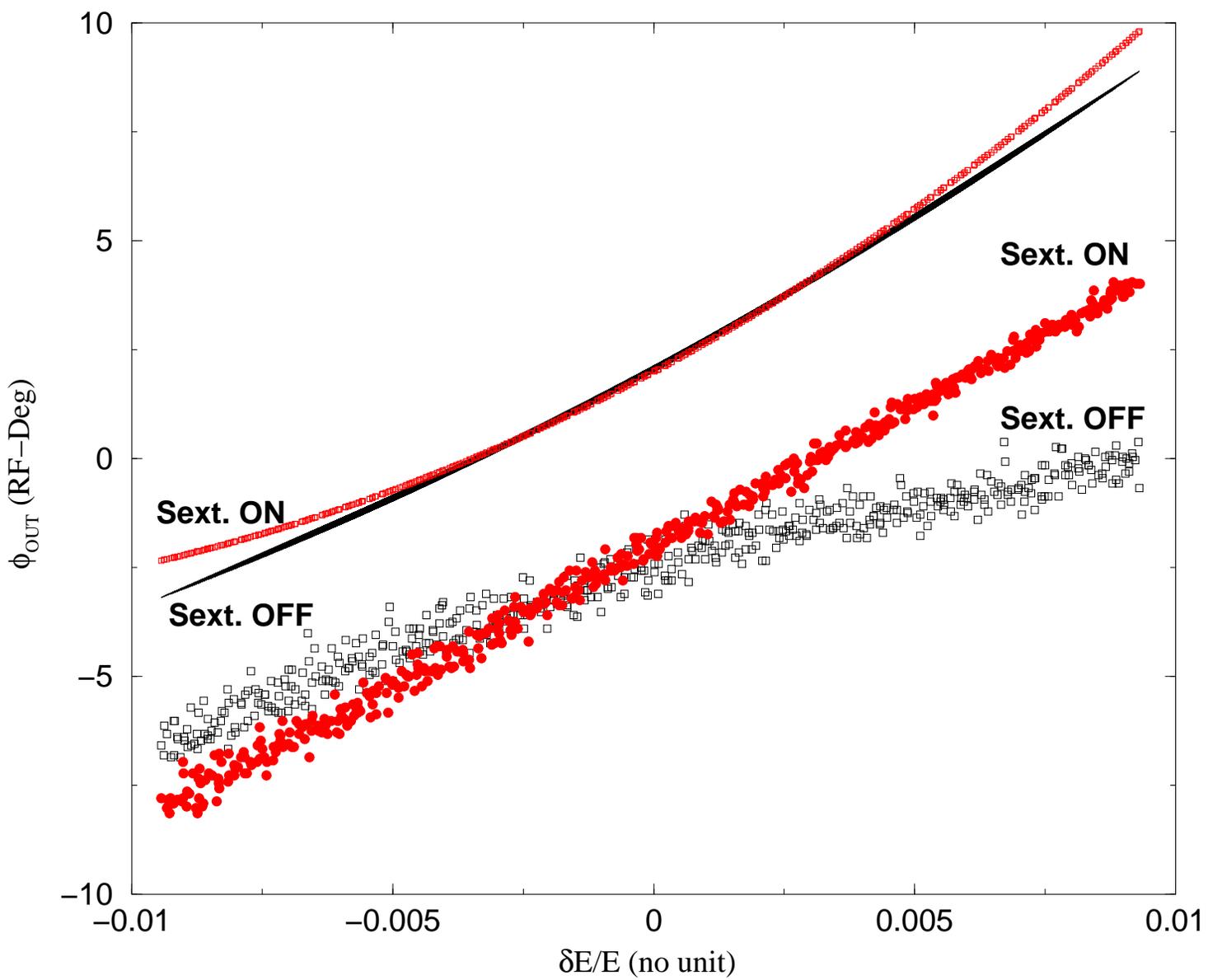}
\caption{\label{fig:sexteffects}Effect of the sextupole of arc~2 on the momentum compaction transfer 
map measured with the detector C3.}
\end{figure}

The sensitivity of the momentum compaction map to second order changes in 
the optics, e.g. imparted by the sextupoles, was also investigated. 
Figure~\ref{fig:sexteffects} depicts the impact of exciting one of the 
sextupole pairs of arc~2 on the map measured by the detector C3. 
The effect observed in simulations and measurements is the same: in both cases 
exciting the sextupole impress a positive curvature on the map -- though 
the agreement is not absolute. This can again be explained by the sensitivity 
of the nonlinear terms to mis-steering via the second order chromatic functions 
$T_{56i}$ (with $i=1,...,4$). Especially when the FEL is operating, the 
aforementioned effect is enhanced because of the large FEL-induced fractional 
momentum spread.

\subsubsection{Other experimental evidence of energy compression:}

The primary evidence of the good performance of the energy compression scheme was our
ability to recover 5~mA average current beam while lasing at high gain with
an average output power of 2.1~kW without any beam losses. 
The Ir-Demo is equipped with a high sensitivity protection system that can 
detect localized losses of beam as low as 1 $\mu$A~\cite{beamlosses}. \\

\begin{figure}
\centering
\leavevmode
\epsfxsize=160mm
\epsfbox{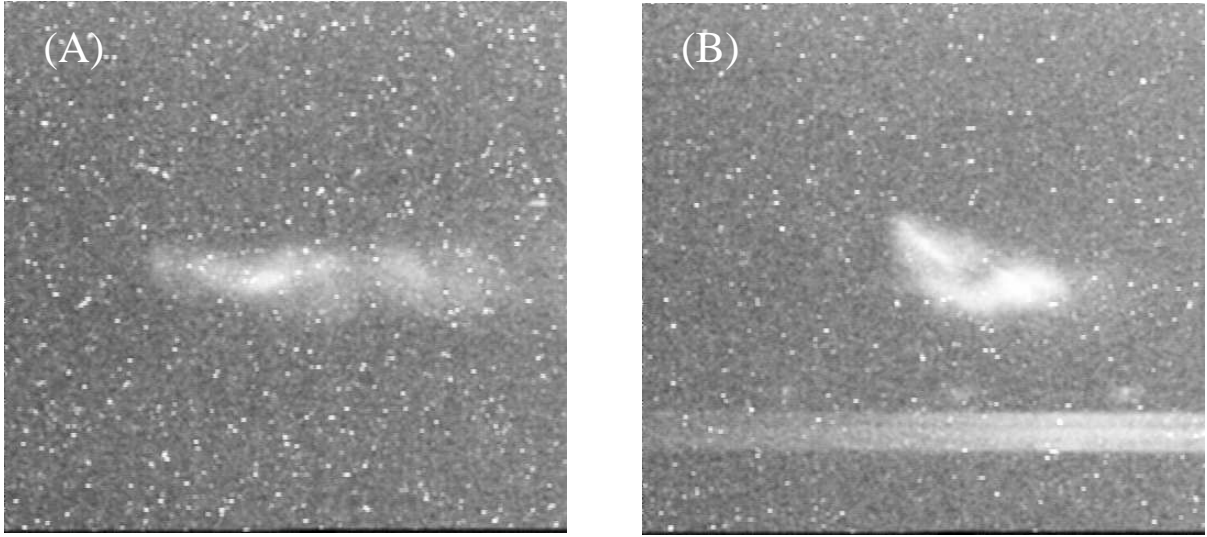}
\caption{Beam density on the energy recovery dump aluminum window observed via 
optical transition radiation. The sextupoles in the arc~2 are turned off 
{\bf (A)}, and excited to their nominal value {\bf (B)}.\label{fig:sexteffectsspot}}
\end{figure}

Another validation of the method is to observe the beam transverse density 
on the energy recovery dump aluminum window. Such a measurement was performed 
by detecting the backward optical transition radiation (OTR) emitted at the 
interface vacuum/aluminum of the window. The OTR images are detected with a 
charge couple device (CCD) camera and digitized for analysis. At the window 
the dispersion (horizontal) is $\eta\simeq 1$~m. The results of our observations are 
presented in Figure~\ref{fig:sexteffectsspot}:
when the longitudinal compression is properly tuned the beam is
tightly focused whereas by slightly mis-setting one pair of sextupoles
(which does not affect \( - \) to first order \( - \) the lattice functions),
the beam horizontal size starts to blow-up, indicating relatively poor
energy spread.

\section{Conclusion}

We have successfully characterized the energy compression scheme to recover 
the ``spent" electron beam after the FEL-process in the IR-demo. Such
techniques are well adapted for energy-recovering an electron beam in a 
moderate power FEL. For very high power FELs anticipated in the future,
higher 
order corrections and/or the use of a dedicated ``accelerating'' section 
to impart the required position energy correlation might be needed. 
As an example the forthcoming upgrade of the Ir-Demo to 10~kW requires the 
use of octupoles~\cite{douglas-10kW-linac2000}. \\

The techniques we have developed for characterizing the longitudinal lattice 
maps have proven to be a valuable tool both during the commissioning of 
the Ir-Demo FEL but also for its day-to-day operation: under the nominal 
setup of the recirculator the map pattern is well defined and changes of this 
pattern indicate one of the component in the recirculator or RF-system is 
not properly set. However there is room for improvement: we have shown that 
in some cases, e.g. due to mis-steering of the beam in the arcs, the extraction 
of quantitative information from the transfer map may turn out to be difficult. 
A way of improving this situation is to measure the transfer map for different 
steering conditions. By doing such a ``two dimensional difference orbit" 
measurement, one could get information on the coupling terms (e.g. $T_{55j}$ 
or $T_{56j}$) and thus correct accordingly the measurement to be less 
sensitive to misalignment. 

\section{Acknowledgments}
This work was sponsored by US-DOE grant number DE-AC05-84ER40150, the Office of 
Naval Research, the Commonwealth of Virginia and the Laser Processing Consortium.

\newpage
\begin{table}[h!]
\begin{center}
\begin{tabular}{c c c}
\hline \hline 
Detector      &  Linear Coeff.  & Quadratic Coeff.  \\ \hline 
{\bf Experiment} \\
C1    &  0.12   & 8$\times 10^{-4}$ \\
C2    & -0.08   & 16$\times 10^{-4}$ \\
C3    &  0.09   & 6$\times 10^{-4}$ \\ \hline
{\bf Simulation} \\
C1    &  0.11   & 7$\times 10^{-4}$ \\
C2    & -0.08   & 3$\times 10^{-4}$ \\
C3    &  0.03   & 4$\times 10^{-4}$ \\ \hline \hline 
\end{tabular}
\caption{Measured and simulated linear and quadratic expansions of the 
compression efficiency map at the three detector location (C1, C2, and C3). 
\label{tab:m55t555}}
\end{center}
\end{table} 

\newpage

\clearpage

\newpage

\newpage


\newpage

\newpage

\newpage

\newpage

\newpage
\begin{figure}
\centering
\leavevmode
\epsfxsize=160mm
\epsfbox{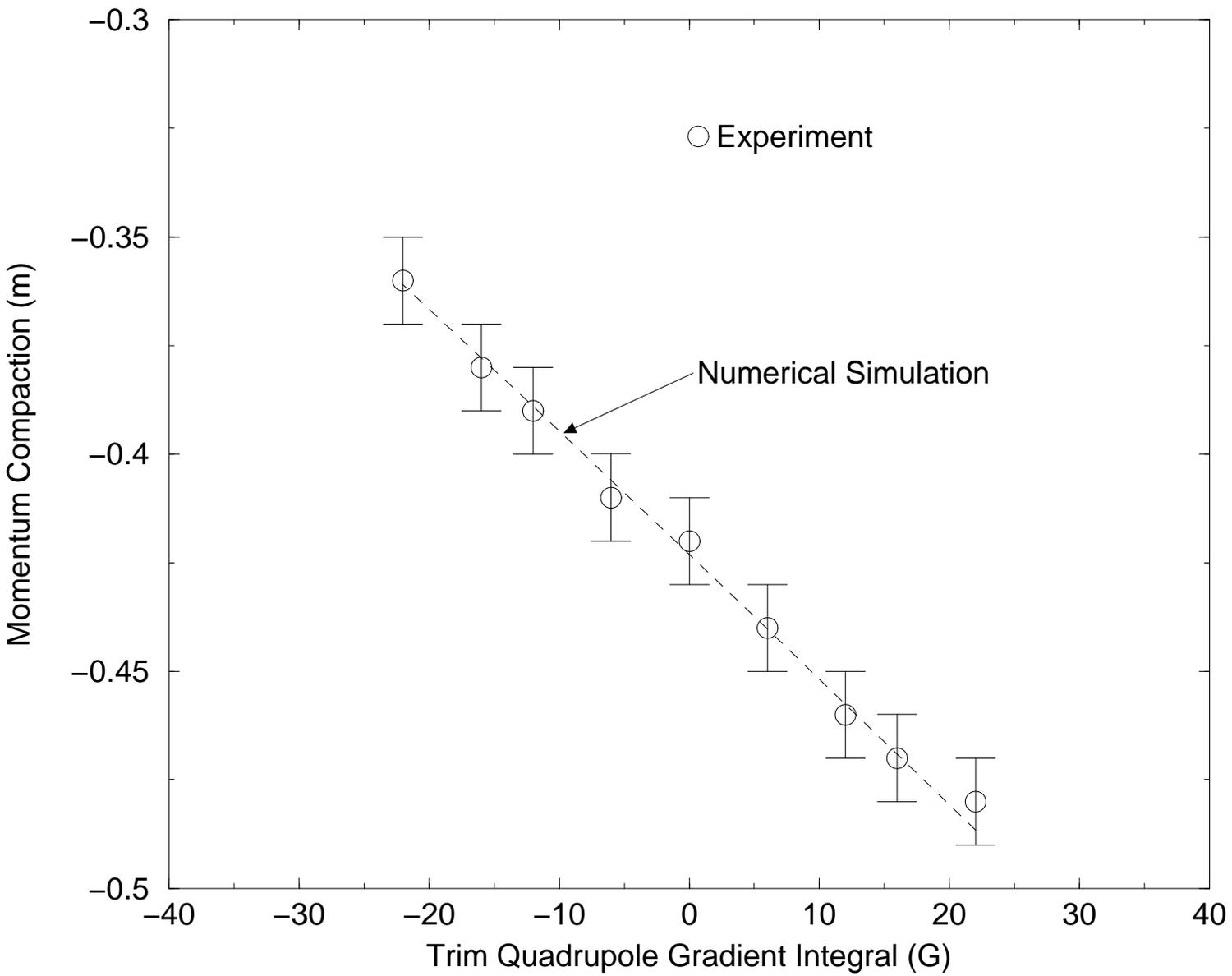}
\caption{Comparison between the expected and measured linear momentum compaction, 
$R_{56}$, for the whole recirculation transport, versus different settings for 
one of the trim quadrupole pairs in one of the 180-deg arc.\label{fig:r56-comparison}}
\end{figure}

\newpage

\newpage

\newpage

\newpage

\end{document}